\documentclass[]{aa}

\usepackage{txfonts}

\usepackage{amsmath}
\usepackage{amssymb,amsfonts,textcomp}

\usepackage{afterpage}

\usepackage{savesym}
\savesymbol{tablenum}
\usepackage{siunitx}
\restoresymbol{SIX}{tablenum}

\usepackage{algorithm}
\usepackage[noend]{algpseudocode}
\algnewcommand{\LineComment}[1]{\State \(\triangleright\) #1}
\usepackage{varwidth}

\DeclareMathOperator*{\argmin}{arg\,min}
\DeclareMathOperator*{\argmax}{arg\,max}

\begin{document}

\title{Image Domain Gridding: a fast method for convolutional resampling of visibilities}
\titlerunning{Image Domain Gridding}

\author{Sebastiaan van der Tol \and Bram Veenboer \and Andr\'e R. Offringa}

\institute{Netherlands Institute for Radio Astronomy (ASTRON), Postbus 2, 7990 AA Dwingeloo, The Netherlands}

\date{Received 20 February 2018 / Accepted 21 March 2018 }


\abstract{
In radio astronomy obtaining a high dynamic range in synthesis imaging of wide fields requires a correction for time and direction-dependent effects.
Applying direction-dependent correction can be done by either partitioning the image in facets
and applying a direction-independent correction per facet, or by including the correction in the gridding kernel (AW-projection).

An advantage of AW-projection over faceting is that the effectively applied beam is a sinc interpolation of the sampled beam, where
the correction applied in the faceting approach is a discontinuous piece wise constant beam.
However, AW-projection quickly becomes prohibitively expensive when the corrections vary over short time scales.
This occurs for example when ionospheric effects are included in the correction.
The cost of the frequent recomputation of the oversampled convolution kernels then dominates the total cost of gridding.

Image domain gridding is a new approach that avoids the costly step of computing oversampled convolution kernels.
Instead low-resolution images are made directly for small groups of visibilities which are then transformed and added to the large  $uv$ grid.
The computations have a simple, highly parallel structure that maps very well onto massively parallel hardware such as graphical processing units (GPUs).
Despite being more expensive in pure computation count, the throughput is comparable to classical W-projection.
The accuracy is close to classical gridding with a continuous convolution kernel.
Compared to gridding methods that use a sampled convolution function, the new method is more accurate.
Hence the new method is at least as fast and accurate as classical W-projection, while
allowing for the correction for quickly varying direction-dependent effects.
}

\maketitle

\section{Introduction}

In aperture synthesis radio astronomy an image of the sky brightness distribution is reconstructed from measured visibilities.
A visibility is the correlation coefficient between the electric field at two different locations.
The relationship between the sky brightness distribution and the expected visibilities is a linear equation
commonly referred to as the `measurement equation' (ME) \citep{Smirnov2011}.

An image could be reconstructed using generic solving techniques, but the computational cost of any reasonably sized problem is prohibitively large.
The cost can be greatly reduced by using the fact that under certain conditions the ME can be approximated by a two-dimensional (2D) Fourier transform.
The discretized version of the ME can then be evaluated using the very efficient fast Fourier transform (FFT).

To use the FFT, the data needs to be on a regular grid. Since the measurements have continuous coordinates, they first need to be resampled onto a regular grid.
In \cite{Brouw1975} a convolutional resampling method is introduced known as ``gridding''.
The reverse step, needed to compute model visibilities on continuous coordinates from a discrete model, is known as ``degridding''.

For larger fields of view the approximation of the ME by a Fourier transform is inaccurate. The reduction of the full three-dimensional (3D) description to two dimensions
only holds when all antennas are in a plane that is parallel to the image plane. Also, the variations of the instrumental and atmospheric effects over the field of view are not included.

There are two approaches to the problem of wide field imaging: 1) Partition the image into smaller sub-images or facets such that the approximations hold for each of the facets.
The facets are then combined together whereby special care needs to be taken to avoid edge effects \citep{Cornwell1992, Tasse2018}; and 2) include deviations from the Fourier transform in the convolution function.
The W-projection algorithm \citep{Cornwell2005} includes the non-coplanar baseline effect.
The A-projection algorithm \citep{Bhatnagar2008} extended upon this by also including instrumental effects.
For the Low-Frequency Array (LOFAR) it is necessary to include ionospheric effects as well \citep{Tasse2013}.
Each successive refinement requires the computation of more convolution kernels.
The computation of the kernels can dominate the total cost of gridding, especially when atmospheric effects are included in the convolution kernel,
because these effects can vary over short time scales.

The high cost of computing the convolution kernels is the main motivation for the development of a new algorithm for gridding and degridding.
The new algorithm presented in this paper effectively performs the same operation as classical gridding and degridding with AW-projection, except that it does this more efficiently
by avoiding the computation of convolution kernels altogether.
Unlike, for example, the approach by \cite{Young2015}, the corrections do not need to be decomposable in a small number of basis functions.

The performance in terms of speed of various implementations of the algorithm on different types of hardware is the subject of \cite{Veenboer2017}.
The focus of this paper is on the derivation of the algorithm and analysis of its accuracy.

The paper is structured as follows:
In section \ref{sec:gridding} we review the gridding method and AW-projection.
In section \ref{sec:imagedomaingridding} we introduce the new algorithm which takes the gridding operation to the image domain.
In section \ref{sec:analysis} the optimal taper for the image domain gridding is derived.
Image domain gridding with this taper results in a lower error than classical gridding with the classical optimal window.
In section \ref{sec:simulations} both the throughput and the accuracy are measured.

The following notation is used throughout the paper.
Complex conjugation of $x$ is denoted $x^{*}$.
Vectors are indicated by bold lower case symbols, for example, $\mathbf{v}$, matrices by bold upper case symbols, $\mathbf{M}$.
The Hermitian transpose of a vector or matrix is denoted $\mathbf{v}^\mathrm{H}$, $\mathbf{M}^\mathrm{H}$ , respectively.
For continuous and discrete (sampled) representations of the same object, a single symbol is used. Where necessary, the discrete version is distinguished
from the continuous one by a superscript indicating the size of the grid, that is, $V^{L\times L}$ is a grid of $L\times L$ pixels sampling continuous function $V$.
Square brackets are used to address pixels in a discrete grid, for example, $V[i,j]$, while parentheses are used for the value at continuous coordinates,  $V(u,v)$.
A convolution is denoted by $\ast$; the (discrete) circular convolution by $\circledast$. The Fourier transform, both continuous and discrete, is denoted by $\mathcal{F}$.
In algorithms we use $\gets$ for assignment.

A national patent (The Netherlands only) for the method presented in this paper has been registered at the European Patent Office in The Hague, The Netherlands \citep{vandertol2017}.
No international patent application will be filed.
Parts of the description of the method and corresponding figures are taken from the patent application.
The software has been released \citep{Veenboer2017-2} under the GNU General Public License (GNU GPL \url{https://www.gnu.org/licenses/gpl-3.0.html}).
The GNU GPL grants a license to the patent for usage of this software and derivatives published under the GNU GPL.
To obtain a license for uses other than under GPL, please contact Astron at secretaryrd@astron.nl.

\begin{figure}[tbp!]
  \includegraphics[width=.45\textwidth]{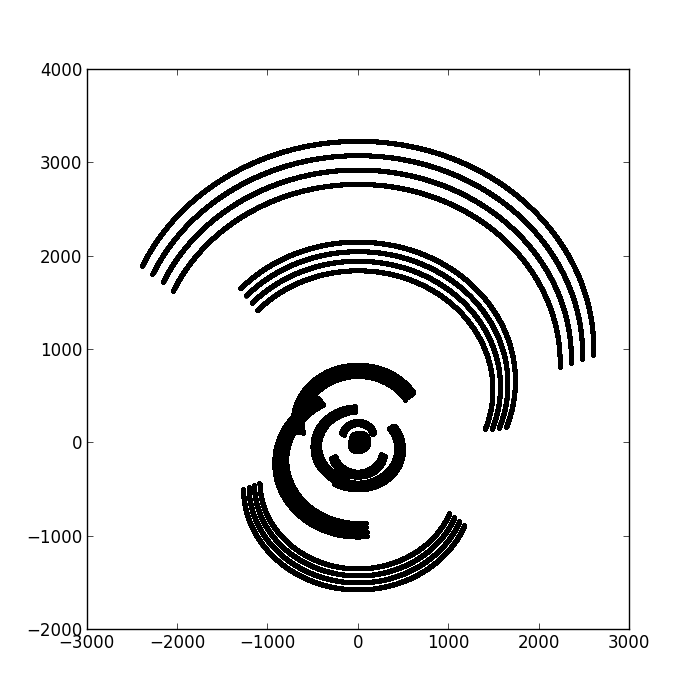}
  \caption{Plot of the  $uv$ coverage of a small subset of an observation. Parallel tracks are for the same baseline, but different for frequencies.}
  \label{fig:uvtrack}
\end{figure}

\section{Gridding}

\label{sec:gridding}

In this section we summarize the classical gridding method. The equations presented here are the starting point for the derivation of image domain gridding in the following section.

The output of the correlator of an aperture synthesis radio telescope is described by the ME \citep{Smirnov2011}.
The full polarization equation can be written as a series of 4x4 matrix products \citep{Hamaker1996} or a series of 2x2 matrix products from two sides \citep{Hamaker2000}.
For convenience, but without loss of generality, the derivations in this paper are done for the scalar (non-polarized) version of the ME.
The extension of the results in this paper to the polarized case is straightforward, by writing out the matrix multiplications in the polarized ME as sums of scalar multiplications.

The scalar equation for visibility $y_{ijqr}$ for baseline $i,j$, channel $q$ at timestep $r$ is given by
\begin{align}
  y_{ijqr} = \iint_{lm} & e^{-j2\pi\left(u_{ijr} l + v_{ijr} m + w_{ijr} n \right)/\lambda_q} \nonumber \\
                        & g_{iqr}(l,m) g_{iqr}^{*} (l,m) I(l,m) dl dm, \label{eq:MeasurmentEquation1}
\end{align}
where  $I(l,m)$ is the brightness distribution or sky image, $(u_{ijr}, v_{ijr}, w_{ijr})$ is the baseline coordinate and $(l,m,n)$ is the direction coordinate, with
$n'=n-1=\sqrt{1-l^2-m^2}-1$, $\lambda_q$ is the wavelength for the $q$th channel,
and $g_{iqr}(l,m)$ is the complex gain pattern of the $i$th antenna.
To simplify the notation we lump indices $i,j,q,r$ together into a single index $k$, freeing indices $i,j,q,r$ for other purposes later on.
Defining
\begin{multline}
   A_k(l,m) \triangleq g_{iqr}(l,m) g_{jqr}^{*} (l,m), \\
   u_k \triangleq u_{ijr}/\lambda_q, \quad v_k \triangleq v_{ijr}/\lambda_q, \quad w_k \triangleq w_{ijr}/\lambda_q,
\end{multline}
allows us to write \eqref{eq:MeasurmentEquation1} as
\begin{equation}
  y_k = \iint_{lm} e^{-j2\pi\left(u_k l + v_k m + w_k n \right)} A_k(l,m) I(l,m) dl dm. \label{eq:MeasurementEquation}
\end{equation}

The observed visibilities $\hat{y}_k$ are modeled as the sum of a model visibility $y_k$ and noise $\eta_k$:
\begin{equation}
  \hat{y}_{k} = y_{k} + \eta_{k}
.\end{equation}
The noise $\eta_{k}$ is assumed to be Gaussian, have a mean of  zero, and be independent for different $k$, with variance $\sigma^2_{k}$.

Image reconstruction is finding an estimate of image  $I(l,m)$ from a set of measurements  $\{\hat{y}_{k}\}$.
We loosely follow a previously published treatment of imaging \cite[][Appendix A]{Cornwell2008}.
To reconstruct a digital image of the sky it is modeled as a collection of point sources.
The brightness of the point source at $(l_i, m_j)$ is given by the value of the corresponding pixel $I[i,j]$.
The source positions $l_i, m_j$ are given by
\begin{equation}
   l_i = -S/2 + iS/L, \quad m_j = -S/2 + jS/L,
\end{equation}
where $L$ is the size of one side of the image in pixels, and $S$ the size of the image projected onto the tangent plane.

Discretization of the image leads to a discrete version of the ME, or DME:
\begin{equation}
  y_k = \sum_{i=1}^L\sum_{j=1}^L e^{-j2\pi\left(u_k l_i + v_k m_j + w_k n'_{ij} \right)} g_k(l_i,m_j) I[i,j] \label{eq:DME}
.\end{equation}
This equation can be written more compactly in matrix form, by stacking the pixels $I[i,j]$ in a vector $\mathbf{x}$, the visibilities $y_k$ in a vector $\mathbf{y}$,
and collecting the coefficients $e^{-j2\pi\left(u_k l_i + v_k m_j + w_k n'_{ij} \right)} g_k(l_i,m_j)$ in a matrix $\mathbf{A}$:
\begin{equation}
  \mathbf{y} = \mathbf{A}\mathbf{x}.
\end{equation}
The vector of observed visibilities $\hat{\mathbf{y}}$ is the sum of the vector of model visibilities $\mathbf{y}$ and the noise vector $\boldsymbol{\eta}$.
Because the noise is Gaussian, the optimally reconstructed image $\hat{\mathbf{x}}$ is a least squares fit to the observed data $\hat{\mathbf{y}}$:
\begin{equation}
   \hat{\mathbf{x}} = \argmin_\mathbf{x} \|\mathbf{\Sigma}^{-1/2}(\mathbf{A}\mathbf{x} - \hat{\mathbf{y}}) \|^2 \label{eg:costfunction}
,\end{equation}
where $\mathbf{\Sigma}$ is the noise covariance matrix, assumed to be diagonal, with $\sigma^2_{k}$ on the diagonal.
The solution is well known and given by
\begin{equation}
    \hat{\mathbf{x}} = \left(\mathbf{A}^\mathsf{H}\mathbf{\Sigma}\mathbf{A}\right)^{-1}\mathbf{A}^\mathsf{H}\mathbf{\Sigma}\hat{\mathbf{y}}
.\end{equation}

In practice the matrices are too large to directly evaluate this equation.
Even if it could be computed, the result would be of poor quality, because matrix $\mathbf{A}^\mathsf{H}\mathbf{\Sigma}\mathbf{A}$ is usually ill-conditioned.
Direct inversion is avoided by reconstructing the image in an iterative manner.
Additional constraints and/or a regularization are applied, either explicitly or implicitly.

Most, if not all, of these iterative procedures need the derivative of cost function \eqref{eg:costfunction} to compute the update.
This derivative is given by
\begin{equation}
  \mathbf{A}^{\mathsf{H}}\mathbf{\Sigma}^{-1}\left(\hat{\mathbf{y}} - \mathbf{A}\mathbf{x}\right) \label{eq:derivative}
.\end{equation}
In this equation, the product $\mathbf{A}\mathbf{x}$ can be interpreted as the model visibilities $\mathbf{y}$ for model image $\mathbf{x}$.
The difference then becomes $\hat{\mathbf{y}}-\mathbf{y}$, which can be interpreted as the residual visibilities.
Finally the multiplication of $\hat{\mathbf{y}}$ or $(\hat{\mathbf{y}}-\mathbf{y})$ by $\mathbf{A}^\mathsf{H}\mathbf{\Sigma}^{-1}$ computes
the dirty, or residual image, respectively.
This is equivalent to the Direct Imaging Equation (DIE), in literature often denoted by the misnomer
\footnote{See footnote on p. 128 of \citet{NRAO1999} on why DFT is a misnomer for this equation.}
Direct Fourier Tranform (DFT):
\begin{equation}
  \hat{I}[i,j]  = \sum_{k=0}^{K-1} e^{\mathrm{j}2\pi\left(u_k l_i + v_k m_j + w_k n'_{ij} \right)} g_k^*(l_i,m_j) \gamma_k \hat{y}_k \label{eq:imaging}
,\end{equation}
where $\gamma_k$ is the weight. The weight can be set to $1/\sigma_k^2$ (the entries of the main diagonal of $\mathbf{\Sigma}^{-1}$) for natural weighting, minimizing the noise,
but often other weighting schemes are used, making a trade off between noise and resolution.
Evaluation of the equations above is still expensive. Because the ME is close to a Fourier transform,
the equations can be evaluated far more efficiently by employing the FFT.

To use the FFT, the measurements need to be put on a regular grid by gridding.
Gridding is a (re)sampling operation in the $uv$ domain that causes aliasing in the image domain, and must therefore be preceded by a filtering operation.
The filter is a multiplication by a taper $c(l,m)$ in the image domain, suppressing everything outside the area to be imaged.
This operation is equivalent to a convolution in the $uv$ domain by $C(u,v)$, the Fourier transform of the taper.
Let the continuous representation of the observed visibilities after filtering be given by
\begin{equation}
   \widetilde{V}(u,v) = \sum_{k=0}^{K-1} y_{k} \delta\left(u - u_{k},v-v_{k}\right) \ast C\left(u,v\right) \label{eq:gridding}
.\end{equation}
Now the gridded visibilities are given by:
\begin{equation}
   \widehat{V}^{L \times L}[i,j] = \widetilde{V}(u_i, v_j) \quad \text{for } 0 \le i,j < L.
\end{equation}
The corresponding image is given by:
\begin{equation}
  \widehat{I}^{L \times L} = \mathcal{F}(\widehat{V}^{L \times L} / c^{L \times L}, \label{eq:avg_beam_correction}
\end{equation}
The division by $c^{L \times L}$ in \eqref{eq:avg_beam_correction} is to undo the tapering of the image by the gridding kernel.

The degridding operation can be described by
\begin{equation}
  y_k \gets \left( V(u,v) \ast C_{k}(u,v) \right)(u_k,v_k),
  \label{eq:degridding}
\end{equation}
where $C_k(u,v)$ is the gridding kernel and $V(u,v)$ is the continuous representation of grid $V^{L \times L}$:
\begin{equation}
  V(u,v) = \sum_{q=0}^{L}\sum_{r=0}^{L} \delta(u-u_r, v-v_r) V[q,r],
\end{equation}
Grid $V^{L \times L}$ is the discrete Fourier transform of model image $I^{L \times L}$ scaled by $\tilde{c}^{L \times L}$:
\begin{equation}
  V \gets \mathcal{F}(I^{L \times L}/\tilde{c}^{L \times L})
\end{equation}

In this form, the reduction in computation cost by the transformation to the $uv$ domain is not immediately apparent.
However, the support of the gridding kernel is rather small, making the equations sparse, and hence cheap to evaluate.
The kernel is the Fourier transform of the window function $c_k(l,m)$.
In the simplest case the window is a taper independent of time index $k$, $c_k(l,m) = b(l,m)$.

The convolution by the kernel in the $uv$ domain applies a multiplication by the window in the image domain.
This suppresses the side-lobes but also affects the main lobe.
A well-behaved window goes towards zero near the edges. At the edges, the correction is unstable and that
part of the image must be discarded. The image needs to be somewhat larger than the region of interest.

The cost of evaluating Eqs. \eqref{eq:gridding} and \eqref{eq:degridding} is determined by the support and the cost of evaluating $C_k$.
The support is the size of the region for which $C_k$ is non-negligible.
Often $C_k$ is precomputed on an over-sampled grid, because then only lookups are needed while gridding.
In some cases $C_k$ can be evaluated directly, but often only an expression in the image domain for $c_k$ is given.
The convolution functions are then computed by evaluating the window functions on a grid that samples the combined image domain effect at least at the Nyquist rate, that is,
the number of pixels $M$, must be at least as large as the support of the convolution function:
\begin{equation}
   c^{M\times M}[i,j] = c(l_i, m_j) \quad \text{for } 0 \leq i,j < M
.\end{equation}
This grid is then zero padded by the oversampling factor $N$ to the number of pixels of the oversampled convolution function $MN \times MN$:
\begin{equation}
C^{MN \times MN} = \mathcal{F}(\mathcal{Z}^{MN \times MN}({c^{M \times M}}))
,\end{equation}
where $l_i = -S/(2M) + iS/M, m_j = -S/(2M) + jS/M$, and $\mathcal{Z}^{MN}$ is the zero padding operator extending a grid to size $MN \times MN$.

Since a convolution in the $uv$ domain is a multiplication in the image domain, other effects that have the form of a multiplication in the image domain can
be included in the gridding kernel as well.
\subsection{W-projection}
In \cite{Cornwell2005} W-projection is introduced.
This method includes the effect of the non coplanar baselines in the convolution function. The corresponding window function is given by
\begin{equation}
c(l,m) = b(l,m) e^{2\pi \mathrm{j} w n'}.
\end{equation}
This correction depends on a single parameter only, the $w$ coordinate.
The convolution functions for a set of $w$ values can be precomputed, and while gridding the nearest $w$ coordinate is selected.
The size of the W term can become very large which makes W projection expensive.
The size of the W term can be reduced by either W-stacking \citep{Humphreys2011} or W-snapshots \citep{Cornwell2012}.

\subsection{A-projection}
A further refinement was introduced in \cite{Bhatnagar2008}, to include the antenna beam as well:
\begin{equation}
   c(l,m) = b(l,m) e^{2\pi \mathrm{j} w n'} g_p(l,m) g_q^{*}(l,m),
\end{equation}
where $g_p(l,m)$ is the voltage reception pattern of the $p$th antenna.

As long as a convolution function is used to sample many visibilities, the relative cost of computing the convolution function is small.
However, for low-frequency instruments with a wide field of view, both the A term and the W term vary over short time scales.
The computation of the convolution kernels dominates over the actual gridding.
The algorithm presented in the following section is designed to overcome this problem by circumventing the need to compute the kernels altogether.

\begin{figure*}[!tbp]
    \includegraphics[width=.45\textwidth]{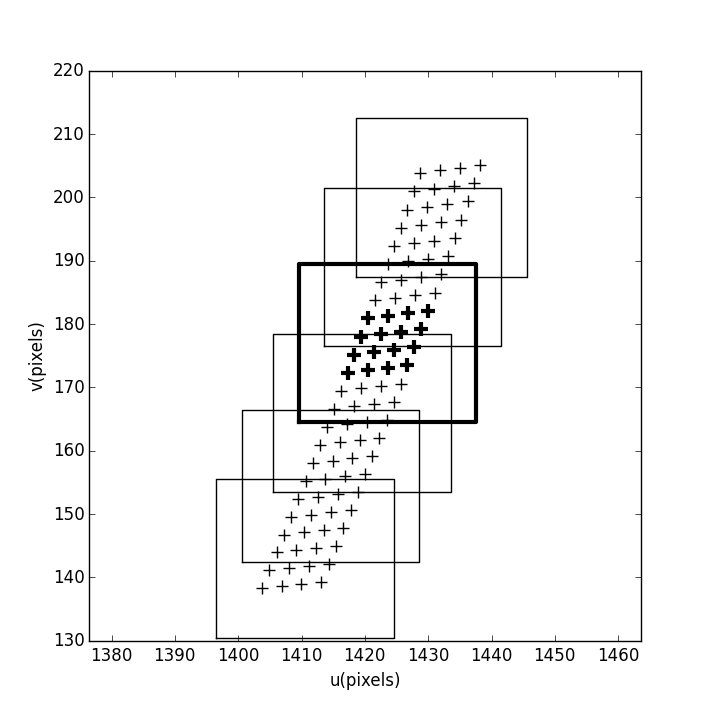}
    \includegraphics[width=.45\textwidth]{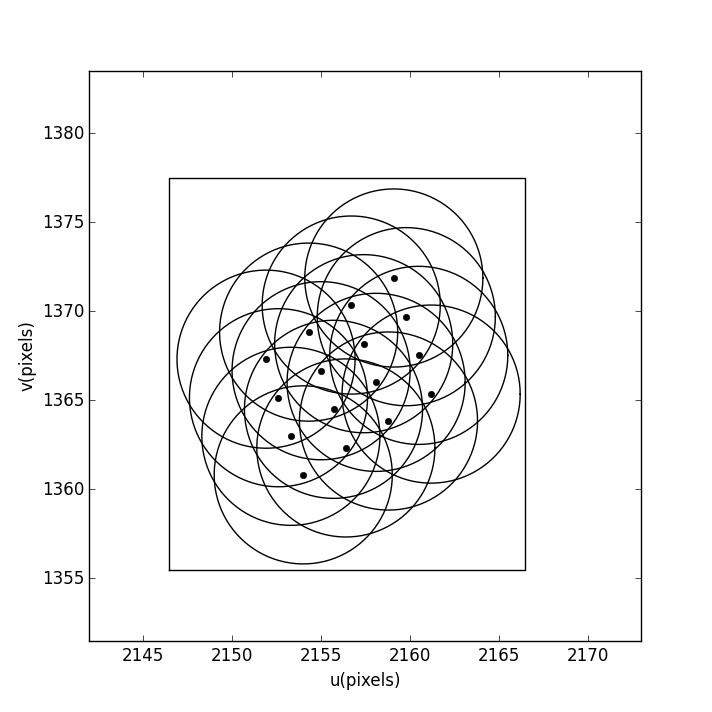}
    \caption{\textit{Left}: Track in $uv$ domain for a single baseline and multiple channels. The boxes indicate the position of the subgrids. The bold box corresponds to the bold samples.
    \textit{Right}: Single subgrid (box) encompassing all affected pixels in the $uv$ grid. The support of the convolution function is indicated by the circles around the samples.
    }
    \label{fig:subgrid}
\end{figure*}

\section{Image domain gridding}

\label{sec:imagedomaingridding}

In this section we present a new method for gridding and degridding. The method is derived from the continuous equations because the results
follow more intuitively than in the discrete form. Discretization introduces some errors, but in the following section the accuracy
of the algorithm is shown to be at least as good as classical gridding.

\subsection{Gridding in the image domain}
Computing the convolution kernels is expensive because they are oversampled.
The kernels need to be oversampled because they need to be shifted to a continuous position in the $uv$ domain.
The key idea behind the new algorithm is to pull part of the gridding operation to the image domain, instead of transforming a zero-padded window function to the $uv$ domain.
In the image domain the continuous $uv$ coordinate of a visibility can be represented by a phase gradient, even if the phase gradient is sampled.
The convolution is replaced by a multiplication of a phase gradient by a window function.
Going back to the image domain seems to defy the reasoning behind processing the data in the $uv$ domain in the first place.
Transforming the entire problem back to the image domain will only bring us back to the original direct imaging problem.

The key to an efficient algorithm is to realize that direct imaging is inefficient for larger images, because of the scaling by the number of pixels.
But for smaller images (in number of pixels) the difference in computational cost between gridding and direct imaging is much smaller.
For very short baselines (small $uv$ coordinates) the full field can be imaged with only a few pixels because the resolution is low.
This can be done fairly efficiently by direct imaging.
Below we introduce a method that makes low-resolution images for the longer baselines too, by partitioning the visibilities first in groups of nearby samples,
and then shifting these groups to the origin of the $uv$ domain. Below we show that these low-resolution images can then be combined in the $uv$
domain to form the final high-resolution image.

\subsection{Partitioning}
The partitioning of the data is done as follows. See Figure \ref{fig:uvtrack} for a typical distribution of data points in the $uv$ domain.
Due to rotation of Earth, the orientation of the antennas changes over time causing the data points to lie along tracks.
Parallel tracks are for observations with the same antenna pair, but at different frequencies.
Figure \ref{fig:subgrid}a shows a close up where the individual data points are visible.
A selection of data points, limited in time and frequency, is highlighted. A tight box per subset around the affected grid points in the (u,v) grid is shown.
The size of the box is determined as shown in Figure \ref{fig:subgrid}b, where the circles indicate the support of the convolution function.

The data is partitioned into  $P$ blocks.
Each block contains data for a single baseline, but multiple timesteps and channels.
The visibilities in the  $p$th group are denoted by $y_{pk} \text{ for } k \in {0,\dots,K_p-1}$, where $K_p$ is the number of visibilities in the block.

The support of the visibilities within a block falls within a box of $L_p \times L_p$ pixels.
We refer to the set of pixels in this box as a subgrid. The position of the central pixel of the  $p$th subgrid is given by  $(u_{0p},v_{0p})$.
The position of the top-left corner of the subgrid in the master grid is denoted by  $(q_{0p}, r_{0p})$.
We note that the visibilities are being partitioned here, not the master $uv$ grid. Subgrids may overlap and the subgrids do not necessarily cover the entire master grid.
\subsection{Gridding equation in the image domain}
A shift from $(u_{0p},v_{0p})$ to the origin can be written as a convolution by the Dirac delta function  $\delta\left(u+u_{0p},v+v_{0p}\right)$.
Partitioning the gridding equation \eqref{eq:gridding} into groups and factoring out the shift for the central pixel leads to
\begin{align}
  \begin{aligned}
    \widehat{V}(u,v) = \sum_{p=1}^M \Big( & \delta\left(u-u_{0p},v-v_{0p}\right) \ast \\
    &
    \begin{aligned}
      & \sum_{k=1}^N
      &
      \begin{aligned} [t]
        & y_{pk} \delta\left(u+u_{0p}, v+v_{0p}\right) \ast \\
        & \Big. \delta\left(u - u_{pk},v-v_{pk}\right) \ast \\
        & C_{pk}\left(u,v\right)\Big).
      \end{aligned}
    \end{aligned}
  \end{aligned}
\end{align}
The shifts in the inner and outer summation cancel each other, leaving only the shift in the original equation \eqref{eq:gridding}.

Now define subgrid  $\widehat{V}_{p}(u,v)$ as the result of the inner summation in the equation above
\begin{align}
    & \widehat{V}_p(u, v) = \nonumber \\
    & \sum_{k=1}^{N} y_{pk}\delta\left(u + u_{0p} - u_{pk} ,v+v_{0p}-v_{pk} \right) \ast C_{pk}\left(u,v\right).
\end{align}
The $uv$ grid $\widehat{V}$ is then a summation of shifted subgrids $\widehat{V}_p$
\begin{align}
    \widehat{V}(u,v) = \sum_{p=1}^M \delta\left(u-u_{0p},v-v_{0p}\right) \ast \widehat{V}_p(u,v).
\end{align}
Now we define the subgrid image $\widehat{I}_p(l,m)$ as the inverse Fourier transform of $\widehat{V}_p$.
The subgrids $\widehat{V}_p$ can then be computed by first computing $\widehat{I}_p(l,m)$ and then transforming it to the $uv$ domain.
The equation for the subgrid image, $\widehat{I}_p(l,m)$, can be found from its definition:
\begin{align}
 \begin{aligned}
 \widehat{I}_p(l,m) & = \mathcal{F}^{-1}\left(\widehat{V}_p(u,v)\right) \\
 &
 \begin{aligned}
 = \sum_{k=1}^{N} & \Big( y_{pk} e^{2\pi\mathrm{i}\left(\left(u_{pk}-u_{0p}\right)l + \left(v_{pk} - v_{0p} \right) m + w_{pk}n\right)} \\
   & c_{pk}\left(l,m\right)\Big)
 \end{aligned}
 \label{eq:ShiftedDirectImaging}
 \end{aligned}.
\end{align}
This equation is very similar to the direct imaging equation \eqref{eq:imaging}. An important difference is the shift towards the origin making the remaining terms  $\left(u_{pk}-u_{0p}\right)$ and
$\left(v_{pk}-v_{0p}\right)$ much smaller than the  $u_k$ and  $v_k$ in the original equation.
That means that the discrete version of this equation can be sampled by far fewer pixels. In fact image  $\widehat{I}_p\left(l,m\right)$ is critically sampled when the number of pixels equals the size of the enclosing box in the $uv$ domain. A denser sampling is not needed since the Fourier transform of a denser sampled image will result in near zero values in the region outside the enclosing box.
The sampled versions of the subgrid and subgrid image are denoted by $\widehat{V}_p[i,j]$  and $\widehat{I}_p[i,j]$ respectively.

Because  $\widehat{I}_p\left(l,m\right)$ can be sampled on a grid with far fewer samples than the original image, it is not particularly expensive to compute
$\widehat {V}_p(u,v)$ by first computing a direct image using \eqref{eq:ShiftedDirectImaging} and then applying the FFT.
The subgrid  $\widehat{V}_p(u,v)$ can then be added to the master grid. The final image $\widehat{I}$ is then the inverse Fourier transform of $\widehat{V}$ divided by the root mean square (rms) window $\overline{c}(l,m)$.

Discretization of the equations above leads to Algorithm \ref{alg:gridding} and \ref{alg:degridding} for gridding and degridding, respectively.
\begin{figure}[tbh]
\begin{algorithm}[H]
  \caption{Image domain gridding}\label{alg:gridding}
  \begin{algorithmic}[0]
    \LineComment{In: \parbox[t]{6cm}{
       $\{y_{pk}\}$ visibilities \\
       $\{u_{pk}\}$, $\{v_{pk}\}$, $\{w_{pk}\}$ : uvw-coordinates \\
       $P, L, \{K_p\}, \{L_p\}$: dimensions \\
       $\{c^{L_p \times L_p}_{pk}\}$ : image domain kernels \\
       $\bar{c}^{L \times L}$: rms image domain kernel}}
    \LineComment{Out: $I^{L \times L}$ image}
    \LineComment{Initialize grid to zero:}
    \State{$V^{L\times L} \gets 0$}
    \LineComment{Iterate over data blocks:}
    \For{p in $0 \ldots P-1$}
      \LineComment{Initialize subgrid to zero:}
      \State{$I_p^{L_p \times L_p} \gets 0$}
      \LineComment{iterate over data within block:}
      \For{k in $0 \ldots K_p-1$}
        \State $u \gets u_{pk} - u_{0p}$
        \State $v \gets v_{pk} - v_{0p}$
        \State $w \gets w_{pk} - w_{0p}$
        \LineComment{iterate over pixels in subgrid:}
        \For{i in $0 \ldots L_p - 1$}
          \For{j in  $0 \ldots L_p - 1$}
            \State $l \gets - \frac{S}{2} + \frac{i}{L_p}\frac{S}{2}$
            \State $m \gets -\frac{S}{2} + \frac{j}{L_p}\frac{S}{2}$
            \State $n' \gets \sqrt{1 - l_q^2 - m_r^2} - 1$
            \State
            \begin{varwidth}[t]{\linewidth}
            $I_p \left[i,j\right] \gets I_p\left[i,j\right] + $ \par
            \hskip\algorithmicindent $e^{2\pi\mathrm{j}\left(ul + vm + wn'\right)} c_{pk}^{*}[i,j] y_{pk}$
            \end{varwidth}
          \EndFor
        \EndFor
      \EndFor
      \LineComment{Transform subgrid to $uv$ domain:}
      \State $V_{p} \leftarrow \operatorname{FFT}(I_p)$
      \LineComment{Add subgrid to master grid:}
      \For{ $i$ in  $0 \ldots L_p - 1$}
        \For{$j$ in  $0 \ldots L_p - 1$}
          \State
            \begin{varwidth}[t]{\linewidth}
              $V\left[i+i_{0p},j + j_{0p}\right] \gets$ \par
              \hskip\algorithmicindent $V\left[i+i_{0p}, j + j_{0p}\right] + V_{p}[i,j]$
            \end{varwidth}
        \EndFor
      \EndFor
    \EndFor
    \LineComment{Transform grid and apply inverse rms taper:}
    \State{$I^{L \times L} \gets FFT(V^{L \times L}) / \bar{c}^{L \times L}$}
  \end{algorithmic}
\end{algorithm}
\end{figure}

\begin{figure}[tbh]
\begin{algorithm}[H]
  \caption{Image domain degridding}\label{alg:degridding}
  \begin{algorithmic}[0]
    \LineComment{In: \parbox[t]{6cm}{
        $I^{L \times L}$ image \\
        $\{u_{pk}\}$, $\{v_{pk}\}$, $\{w_{pk}\}$ : uvw-coordinates \\
        $L, \{K_p\}, \{L_p\}$: dimensions \\
        $\{c^{L_p \times L_p}_{pk}\}$ : image domain kernels \\
        $\bar{c}^{L \times L}$: rms image domain kernel}}
    \LineComment{Out: $\{y_{pk}\}$ visibilities}
    \LineComment{Apply inverse rms taper to entire image:}
    \State $I \gets I/c$
    \LineComment{Fourier transform entire image:}
    \State $V \gets \mathrm{FFT}(I)$
    \LineComment{Iterate over data blocks:}
    \For{p in $0 \ldots P-1$}
        \LineComment{initialize subgrid from master grid:}
        \For{i in  $0\ldots L_p - 1$}
        \For{j in  $0\ldots L_p - 1$}
            \State $V_p[i,j] \gets V[i+i_{0p},j+j_{0p}]$
        \EndFor
        \EndFor
        \LineComment{Transform subgrid to image domain:}
        \State $I_p \gets IFFT(V_p)$

        \For{k in  $0 \ldots K_p-1$}
        \State $\Delta u \gets u_{pk} - u_{0p}$
        \State $\Delta v \gets v_{pk} - v_{0p}$
        \State $\Delta w \gets w_{pk} - w_{0p}$
        \State $y_{pk} \gets 0$
        \For{i in  $0 \ldots L_p - 1$}
            \For{j in  $0 \ldots L_p - 1$}
            \State $l \leftarrow -\frac{S}{2} + \frac{i}{L_p-1}\frac{S}{2}$
            \State $m \leftarrow -\frac{S}{2} + \frac{j}{L_p - 1} \frac{S}{2}$
            \State $n' \leftarrow \sqrt{1-l_i^2-m_j^2} - 1$
            \State
            \begin{varwidth}[t]{\linewidth}
            $y_{pk} \gets  y_{pk} + $ \par
            \hskip\algorithmicindent $e^{-2\pi\mathrm{j} \left(\Delta ul + \Delta vm + \Delta wn'\right)} c_{pk} \left[i,j\right] I \left[ i,j \right]$
            \end{varwidth}
            \EndFor
        \EndFor
        \EndFor
    \EndFor
  \end{algorithmic}
\end{algorithm}
\end{figure}

\subsection{Variations}

The term for the `convolution function' $c_{pk} \left[i,j\right]$ is kept very generic here by giving it an index $k$, allowing a different value for each sample.
Often the gain term, included in $c_{pk}$, can be assumed constant over many data points.
The partitioning of the data can be done such that only a single $c_p$ for each block is needed.
The multiplication by $c_k$ can then be pulled outside the loop over the visibilities, reducing the number of operations in the inner loop.

In the polarized case each antenna consists of two components, each measuring a different polarization, the visibilities are 2$\times$2 matrices and the gain $g$ is described by a 2$\times$2 Jones matrix.
For this case the algorithm is not fundamentally different. Scalar multiplications are substituted by matrix multiplications, effectively adding an extra loop over the different polarizations of the data, and an extra loop over the differently polarized images.

\begin{table*}[tp]
    \caption{Level of aliasing for different gridding methods and kernel sizes}
    \label{tab:error_idg}
    \centering
    \begin{tabular}{cccccccc}
      \hline\hline
      \multicolumn{1}{c}{$\beta$}  & \multicolumn{1}{c}{Classical}  & \multicolumn{6}{c} {Image domain gridding} \\
      \multicolumn{1}{c}{} & \multicolumn{1}{c}{PSWF}  & \multicolumn{1}{c}{$L$=8} & \multicolumn{1}{c}{$L$=16} & \multicolumn{1}{c}{$L$=24} & \multicolumn{1}{c}{$L$=32} & \multicolumn{1}{c}{$L$=48} & \multicolumn{1}{c}{$L$=64} \\
      \hline
      3.0 & 3.33e-02 & 4.63e-02 & \textbf{2.87e-02} & \textbf{2.25e-02} & \textbf{1.92e-02} & \textbf{1.54e-02} & \textbf{1.32e-02} \\
      5.0 & 1.68e-03 & 4.60e-03 & 2.59e-03 & 1.94e-03 & \textbf{1.60e-03} & \textbf{1.25e-03} & \textbf{1.06e-03} \\
      7.0 & 7.96e-05 & 2.99e-04 & 1.67e-04 & 1.25e-04 & 1.03e-04 & 7.89e-05 & \textbf{6.62e-05} \\
      9.0 & 3.68e-06 &  & 9.08e-06 & 6.96e-06 & 5.78e-06 & 4.45e-06 & 3.71e-06 \\
      11.0 & 1.68e-07 &  & 4.82e-07 & 3.55e-07 & 2.98e-07 & 2.33e-07 & 1.95e-07 \\
      13.0 & 7.58e-09 &  & 2.70e-08 & 1.79e-08 & 1.47e-08 & 1.16e-08 & 9.83e-09 \\
      15.0 & 3.40e-10 &  & 1.45e-09 & 9.15e-10 & 7.25e-10 & 5.61e-10 & 4.78e-10 \\
      \hline
    \end{tabular}
    \tablefoot{Level of aliasing for classical gridding with the PSWF and for image domain gridding with the optimal window for different subgrid sizes $L$, and different kernel sizes $\beta$.
    Numbers in bold indicate where image domain gridding has lower aliasing than classical gridding.
    The numbers in this table were generated with code using mpmath (1), a Python library for arbitrary-precision floating-point arithmetic.}
    \tablebib{(1) \citet{mpmath}}
\end{table*}

\section{Analysis}
\label{sec:analysis}

In the previous section, the image domain gridding algorithm is derived rather intuitively without considering the effects of sampling and truncation,
except for the presence of a still-unspecified anti-aliasing window $c[i,j]$.
In this section, the output of the algorithm is analyzed in more detail. The relevant metric here is the difference between the result of direct imaging/evaluation
and gridding/degridding, respectively. This difference, or gridding error, is due solely to the side lobes of the anti-aliasing window.
These side lobes are caused by the limited support of the gridding kernel. An explicit expression for the error will be derived in terms of the anti-aliasing window.
Minimization of this expression leads directly to the optimal window, and corresponding error. This completes the analysis of accuracy of image domain gridding,
except for the effect of limited numerical precision, which was found in practice not to be a limiting factor.

For comparison we summarize the results on the optimal anti-aliasing window for classical gridding known in the literature.
For both classical and image domain gridding the error can be made arbitrarily small by selecting a sufficiently large kernel.
It is shown below that both methods reach comparable performance for equal kernel sizes.
Conversely, to reach a given level of performance both methods need kernels of about the same size.

\subsection{Optimal windows for classical gridding}
We restrict the derivation of the optimal window in this section to a 1D window $f(x)$.
The spatial coordinate is now $x$, replacing the $l,m$ pair in the 2D case, and normalized such that the region to be imaged, or the main lobe, is given by $-1/2 \leq x < 1/2$.
The 2D windows used later on are a simple product of two one-dimensional (1D) windows, $c(l,m) = f(l/S)f(m/S)$, and it is assumed that optimality is mostly preserved.
\citet{Brouw1975} uses as criterion for the optimal window that it maximizes the energy in the main lobe relative to the total energy:
\begin{equation}
  f_{\mathrm{opt}} = \argmax_{f} \frac{\int_{-1/2}^{1/2}\|f(x)\|^2\,\mathrm{d}x}{\int_{-\infty}^{\infty}\|f(x)\|^2\,\mathrm{d}x},
\end{equation}
under the constraint that its support in the $uv$ domain is not larger than a given kernel size $\beta$.
This minimization problem was already known in other contexts. In \cite{Slepian1961} and \cite{Landau1961} it is shown that this problem can be written as an eigenvalue problem. The solution is the prolate spheroidal wave function (PSWF).
The normalized energy in the side lobes is defined by
\begin{equation}
  \varepsilon^2 = \frac{\int_{-\infty}^{-1/2}\|f(x)\|^2\,\mathrm{d}x + \int_{1/2}^{+\infty}\|f(x)\|^2\,\mathrm{d}x}{\int_{-\infty}^{\infty}\|f(x)\|^2\,\mathrm{d}x}.
\end{equation}
For the PSWF, the energy in the side lobes is related to the eigenvalue:
\begin{equation}
   \varepsilon^2_{\textsc{\tiny PSWF}} = 1 - \lambda_0(\alpha),
\end{equation}
where $\lambda_0(\alpha)$ is the first eigenvalue and $\alpha = \beta\pi/2$. The eigenvalue is given by:
\begin{equation}
  \lambda_0(\alpha) = \frac{2\alpha}{\pi}\left[R_{00}(\alpha,1)\right]^2,
\end{equation}
where $R_{mn}(c, \eta)$ is the radial prolate spheroidal wave function \citep[ch. 21]{Abramowitz1964}.

The second column of Table \ref{tab:error_idg} shows the aliasing error $\varepsilon$ for different $\beta$.
The required kernel size can be found by looking up the smallest kernel that meets the desired level of performance.

\subsection{Effective convolution function in image domain gridding}
In classical gridding, the convolution by a kernel in the $uv$ domain effectively applies a window in the image domain.
In image domain gridding, the convolution by a kernel is replaced by a multiplication on a small grid in the image domain
by a discrete taper $c[i,j]$. Effectively this applies a (continuous) window on the (entire) image domain, like in classical gridding.
Again the 2D taper is chosen to be a product of two 1D tapers:
\begin{equation}
   c[i,j] = a_i a_j.
\end{equation}
The 1D taper is described by the set of coefficients $\{a_k\}$.

\begin{figure}
  \includegraphics[width=.48\textwidth]{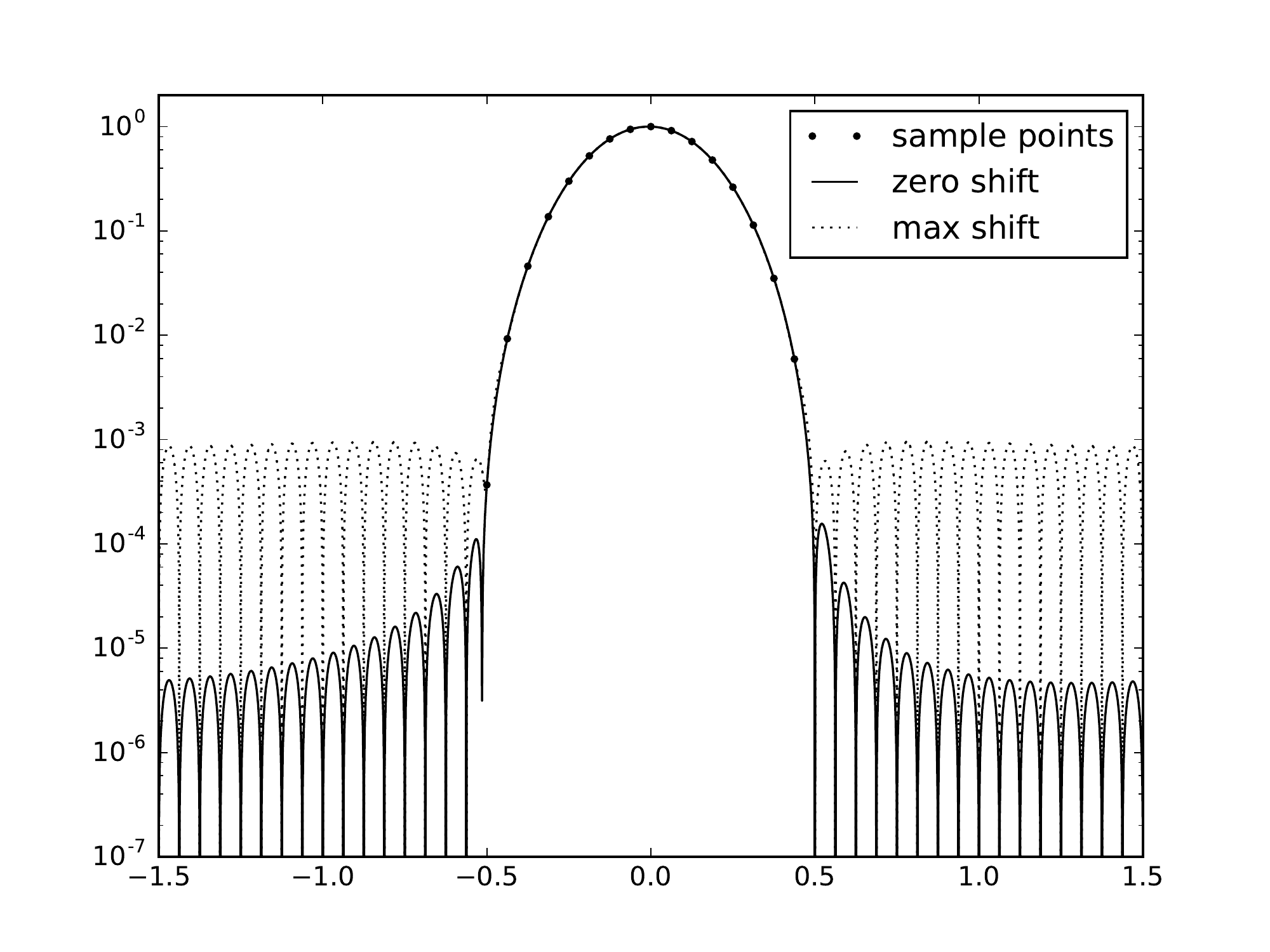}
  \caption{Effective window depending on the position of the sample within the sub-grid. The lowest side lobes are for a sample in the center of the sub-grid.
  The higher side lobes for samples close to the edge are caused by the phase gradient corresponding to a shift away from the center.}
  \label{fig:optimal_window}
\end{figure}

It can be shown that the effective window is a sinc interpolation of the discrete window.
The interpolation however is affected by the multiplication by the phase gradient corresponding to the position shift from
the subgrid center, $\Delta u, \Delta v$. For the 1D analysis, we use a single parameter for the position shift, $s$.

\begin{equation}
  f(x,s) = \sum_{k=0}^{L-1} a_k z_k(s) \operatorname{sinc}(L(x - x_k)) z^{*}(x,s), \label{eq:effective_window}
\end{equation}
where $z(x,s)$ is the phase rotation corresponding to the shift $s$, and $\operatorname{sinc}(x)$ is the normalized sinc function defined by
\begin{equation}
  \operatorname{sinc}(x) = \frac{\sin (\pi x)}{\pi x}.
\end{equation}
Phasor $z(x,s)$ is given by:
\begin{equation}
  z(x,s) = e^\frac{\mathrm{j} 2\pi x s}{L}.
\end{equation}
The sample points are given by $x_k = -1/2 + k/L$. The phasor at the sample points is given by $z_k(s) = z(x_k, s)$.
Although the gradients cancel each other exactly at the sample points, the effect of the gradient can still be seen in the side lobes.
The larger the gradient, the larger the ripples in the sidelobes, as can be seen in Figure \ref{fig:optimal_window}.
Larger gradients correspond to samples further away from the subgrid center.

The application of the effective window can also be represented by a convolution in the $uv$ domain, whereby the kernel depends on the position of the sample within the subgrid.
Figure \ref{fig:convfunc}a shows the convolution kernel for different position shifts. For samples away from the center the convolution kernel is asymmetric.
That is because each sample affects all points in the sub-grid, and not just the surrounding points as in classical gridding.
Samples away from the center have more neighboring samples on one side than the other.
In contrast to classical gridding, the convolution kernel in image domain gridding has side lobes.
These side lobes cover the pixels that fall within the sub-grid, but outside the main lobe of the convolution kernel.

\subsection{Optimal window for image domain gridding}
\begin{figure*}[!t]
    \includegraphics[width=.45\textwidth, trim = 0pt 15pt 0pt 0pt, clip=true]{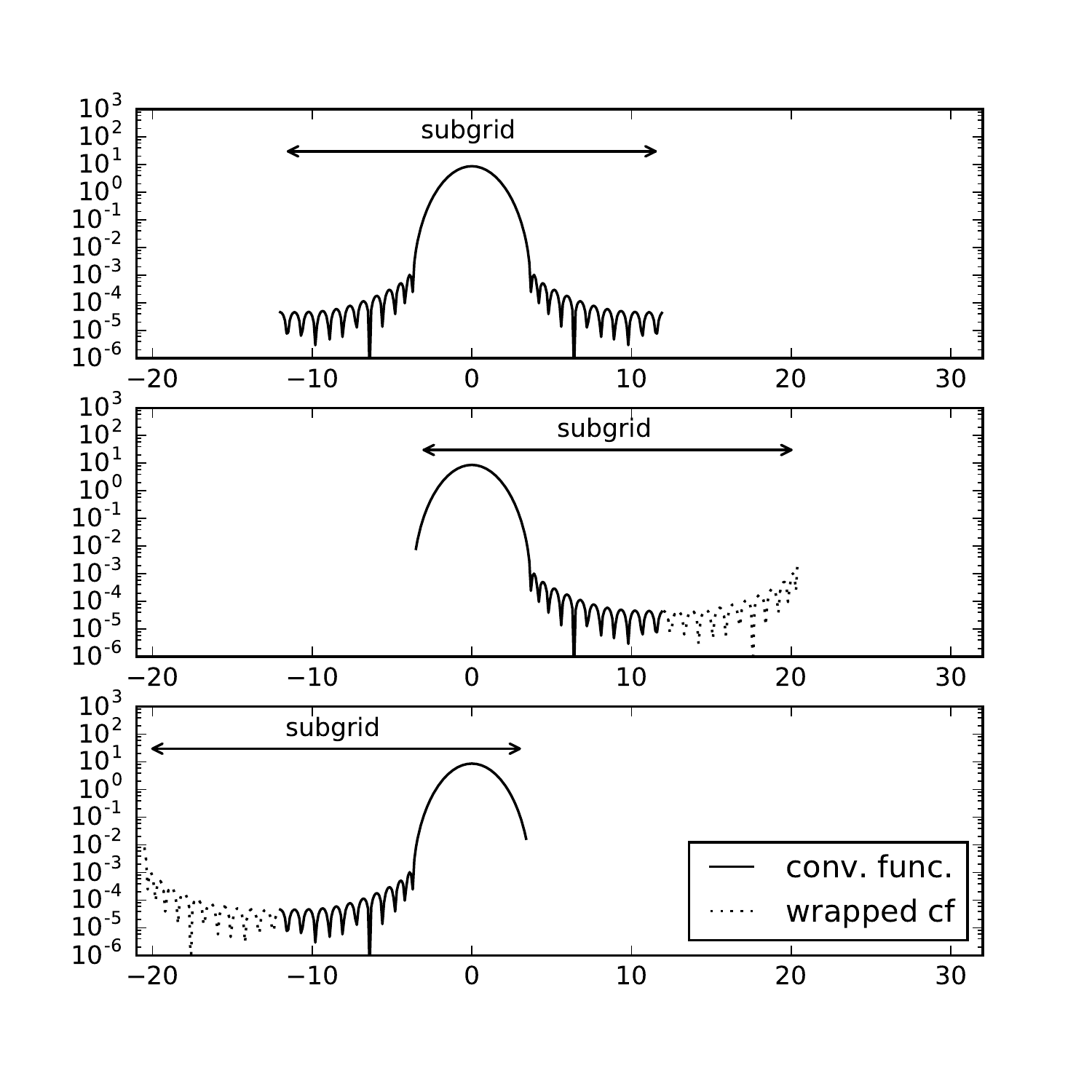}
    \includegraphics[width=.45\textwidth]{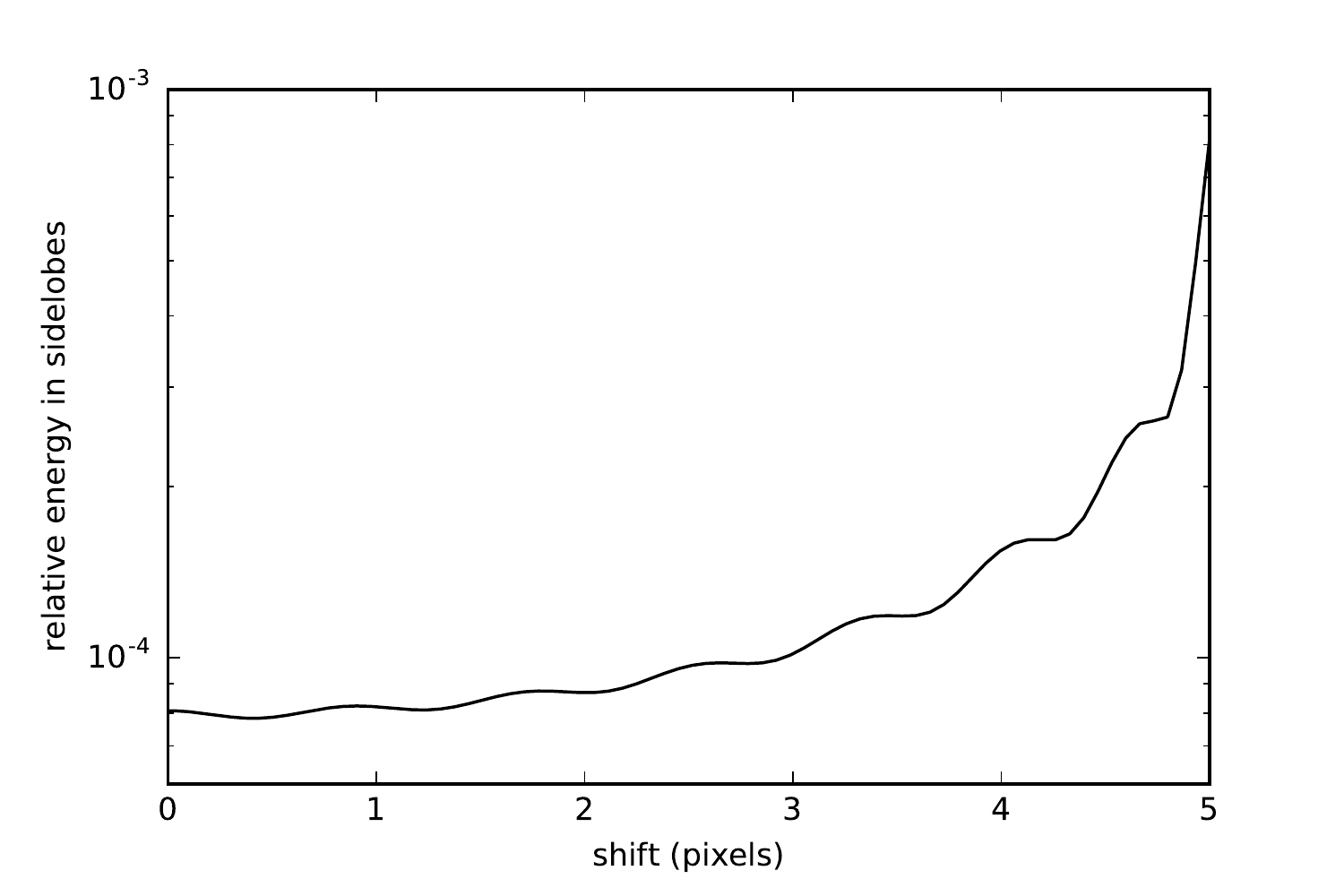}
    \caption{\textit{Left}: Effective convolution function for samples at different positions within the sub-grid;
    \textit{top left}: sample at the center of the sub-grid; \textit{left middle}: sample at the leftmost position within the sub-grid before the main lobe wraps around;
    \textit{left bottom}: sample at the rightmost position within the sub-grid before the main lobe wraps around.
    \textit{Right}: Gridding error as a function of position of the sample within the sub-grid. Close to the center of the sub-grid the error changes little with position.
    The error increases quickly with distance from the center immediately before the maximum distance is reached.}
\label{fig:convfunc}
\end{figure*}

The cost function that is minimized by the optimal window is the mean square of the side lobes of the effective window.
Because the effective window depends on the position within the sub-grid, the mean is also taken over all allowed positions.
For a convolution kernel with main lobe width $\beta$, the shift away from the sub-grid center $\|s\|$ cannot be more than $(L-\beta+1)/2$,
because then the main lobe wraps around far enough to touch the first pixel on the other side.
This effect can be seen in Figure \ref{fig:convfunc}b. The samples in the center have a low error.
The further the sample is from the center, the larger is the part of the convolution kernel that wraps around, and the larger are the side lobes of the effective window.

The cost function to be minimized is given by:

% \begin{equation}
%   \varepsilon^2 = \int_{-(L-\beta-1)/2}^{(L-\beta-1)/2}  \left( \int_{-\infty}^{-0.5} \|f(x,s)\|^2\,\mathrm{d}x +
%   \int_{.5}^{\infty} \|f(x,s)\|^2\,\mathrm{d}x \right )\mathrm{d}s \label{eq:error}
% \end{equation}

\begin{align}
  \varepsilon^2 = \int_{-(L-\beta-1)/2}^{(L-\beta-1)/2}  & \left( \int_{-\infty}^{-0.5} \|f(x,s)\|^2\,\mathrm{d}x \right. + \nonumber \\
  & \left. \int_{.5}^{\infty} \|f(x,s)\|^2\,\mathrm{d}x \right )\mathrm{d}s. \label{eq:error}
\end{align}

In the Appendix a $L \times L$ matrix $\mathbf{\overline{R}}$ is derived such that the error can be written as:
\begin{equation}
  \varepsilon^2 = \mathbf{a}^{H} \mathbf{\overline{R}} \mathbf{a}
,\end{equation}
where $\mathbf{a} = \left[\begin{array}{ccc} a_0 \dots a_{L-1} \end{array}\right]$ is a vector containing the window's coefficients.
The minimization problem:
\begin{equation}
  \mathbf{a}_{opt} = \argmin_{\mathbf{a}} \varepsilon^2(\mathbf{a}) = \argmin_{\mathbf{a}}  \mathbf{a}^{H} \mathbf{\overline{R}} \mathbf{a}
  \label{eq:a_opt}
,\end{equation}
can be solved by a eigenvalue decomposition of $\mathbf{\overline{R,}}$
\begin{equation}
  \mathbf{\overline{R}} = \mathbf{U} \mathbf{\Lambda} \mathbf{U}^{\mathrm{H}}
.\end{equation}
The smallest eigenvalue $\lambda_{L-1}$ gives the side lobe level $\varepsilon$ for the optimal window $\mathbf{a}_{opt} = \mathbf{u}_{L-1}$.

The shape of the convolution kernel is a consequence of computing the cost function as the mean over a range of allowed shifts $-(L-\beta+1) \leq s \leq L-\beta+1$.
The minimization of the cost function leads to a convolution kernel with a main lobe that is approximately $\beta$ pixels wide, but this width is not enforced by any other means than through the cost function.

Table \ref{tab:error_idg} shows the error level for various combinations of sub-grid size $L$ and width $\beta$.
For smaller sub-grids the aliasing for image domain gridding is somewhat higher than for classical gridding with the PSWF,
but, perhaps surprisingly, for larger sub-grids the aliasing is lower.
This is not in contradiction with the PSWF being the convolution function with the lowest aliasing for a given support size.
The size of the effective convolution function of image domain gridding is $L$, the width of the sub-grid, even though the main lobe has only size $\beta$.
Apparently the side lobes of the convolution function contribute a little to alias suppression.

In the end, the exact error level is of little importance. One can select a kernel size that meets the desired performance.
Table \ref{tab:error_idg} shows that kernel size in image domain gridding will not differ much from the kernel size required in classical gridding.
For a given kernel size there exists a straightforward method to compute the window.
In practice the kernel for classical gridding is often sampled. In that case, the actual error is larger than derived here.
Image domain gridding does not need a sampled kernel and the error level derived here is an accurate measure for the level reached in practice.

\section{Application to simulated and observed data}
\label{sec:simulations}

In the previous section it was shown that by proper choice of the tapering window the accuracy of image domain gridding is at least as good as classical gridding.
The accuracy at the level of individual samples was measured based on the root mean square value of the side lobes of the effective window.
In practice, images are made by integration of very large datasets.
In this section we demonstrate the validity of the image domain gridding approach by applying the algorithm in a realistic scenario to
both simulated and observed data, and comparing the result to the result obtained using classical gridding.

\subsection{Setup}
The dataset used is part of a LOFAR observation of the "Toothbrush" galaxy cluster by \cite{vanWeeren2017}.
For the simulations this dataset was used as a template to generate visibilities with the same metadata as the preprocessed visibilities in the dataset.
The pre-imaging processing steps of flagging, calibration and averaging in time and frequency had already been performed.
The dataset covers ten LOFAR sub-bands whereby each sub-band is averaged down to 2 channels, resulting in 20 channels covering the frequency range 130-132 MHz.
The observation included 55 stations, where the shortest baseline is 1km, and the longest is 84km.
In time, the data was averaged to intervals of \SI{10}{seconds}.
The observation lasted \SI{8.5}{hours}, resulting in 3122 timesteps, and, excluding autocorrelations, 4636170 rows in total.

The imager used for the simulation is a modified version of WSClean \citep{Offringa2014}.
The modifications allow the usage of the implementation of image domain gridding by \cite{Veenboer2017} instead of classical gridding.

\subsection{Performance metrics}
Obtaining high-quality radio astronomical images requires deconvolution.
Deconvolution is an iterative process. Some steps in the deconvolution cycle are approximations.
Not all errors thus introduced necessarily limit the final accuracy that can be obtained.
In each following iteration,  the approximations in the previous iterations can be corrected for.
The computation of the residual image however is critical.
If the image exactly models the sky then the residual image should be noise only, or zero in the absence of noise.
Any deviation from zero sets a hard limit on the attainable dynamic range.

The dynamic range can also be limited by the contribution of sources outside the field of view.
Deconvolution will not remove this contribution. The outside sources show up in the image
through side lobes of the point spread function (PSF) around the actual source, and as alias inside the image.
The aliases are suppressed by the anti-aliasing taper.

The PSF is mainly determined by the $uv$ coverage and the weighting scheme, but the gridding method has some effect too.
A well behaved PSF allows deeper cleaning per major cycle, reducing the number of major cycles.

The considerations above led to the following metrics for evaluation of the image domain gridding algorithm
\begin{enumerate}
\item level of the side lobes of the PSF;
\item root mean square level of the residual image of a simulated point source, where the model image and the model to generate the visibilities are an exact match;
\item the rms level of a dirty image of a simulated source outside the imaged area.
\end{enumerate}

\subsection{Simulations}

The simulation was set up as follows.
An empty image of 2048$\times$2048 pixels was generated with cell size of \SI{1}{arcsec}.
A single pixel at position (1000,1200) in this image was set to \SI{1.0}{Jy}.

The visibilities for this image were computed using three different methods: 1) Direct evaluation of the ME, 2) classical degridding, and 3) image domain degridding.
For classical gridding we used the default WSClean settings: a Kaiser-Bessel (KB) window of width 7 and oversampling factor 63.
The KB window is easier to compute than the PSWF, but its performance is practically the same.
For image domain gridding a rather large sub-grid size of 48 $\times$ 48 pixels was chosen.
A smaller sub-grid size could have been used if the channels had been partitioned into groups, but this was not yet implemented.

The image domain gridder ran on a NVIDIA GeForce 840M, a GPU card for laptops.
The CPU is a dual core Intel i7 (with hyperthreading) running at \SI{2.60}{\giga \hertz} clockspeed.

The runtime is measured in two ways: 1) At the lowest level, purely the (de)gridding operation and 2) at the highest level, including all overhead.
The low-level gridding routines report their runtime and throughput. WSClean reports the time spend in gridding, degridding, and deconvolution.
The gridding and degridding times reported by WSClean include the time spent in the large-scale FFTs and reading and writing the data and any other overhead.

The speed reported by the gridding routine was \SI{4.3}{\mega visibilities \per \second}.
For the $20 \times 4636170$ = \SI{93}{\mega visibilities} in the dataset, the gridding time is \SI{22}{\second}.
The total gridding time reported by WSClean was \SI{72}{\second}.

The total runtime for classical gridding was \SI{52}{\second} for Stokes I only, and \SI{192}{\second} for all polarizations.
The image domain gridder always computes all four polarizations.

\begin{figure}[!t]
  \includegraphics[width=0.45\textwidth]{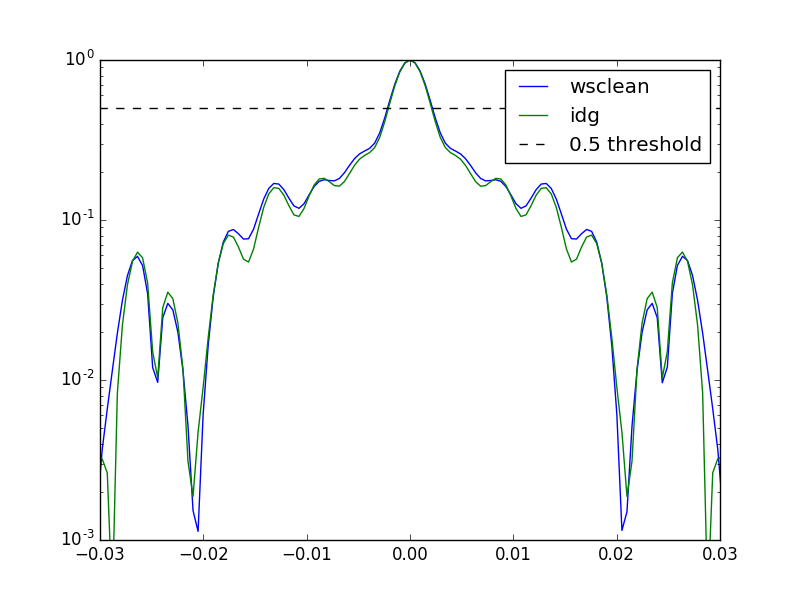}
  \caption{PSF for the classical gridder (blue) and the image domain gridder (green) on a logarithmic scale.
   The main lobes are practically identical. The first side lobes are a bit less for the image domain gridder.
   There are some differences in the further (lower) sidelobes as well, but without a consistent pattern.
   The rms value over the entire image, except the main lobe, is about 5\% lower for image domain gridding than for classical gridding.}
  \label{fig:psf}
\end{figure}

Figure \ref{fig:psf} shows the PSF. In the main lobe the difference between the two methods is small. The side lobes
for the image domain gridder are somewhat (5 \%) lower than for classical gridder. Although in theory this affects the cleaning depth per major cycle, we do not expect
such a small difference to have a noticable impact on the convergence and total runtime of the deconvolution.

\begin{figure*}[!tbp]
  \includegraphics[width=0.33\textwidth]{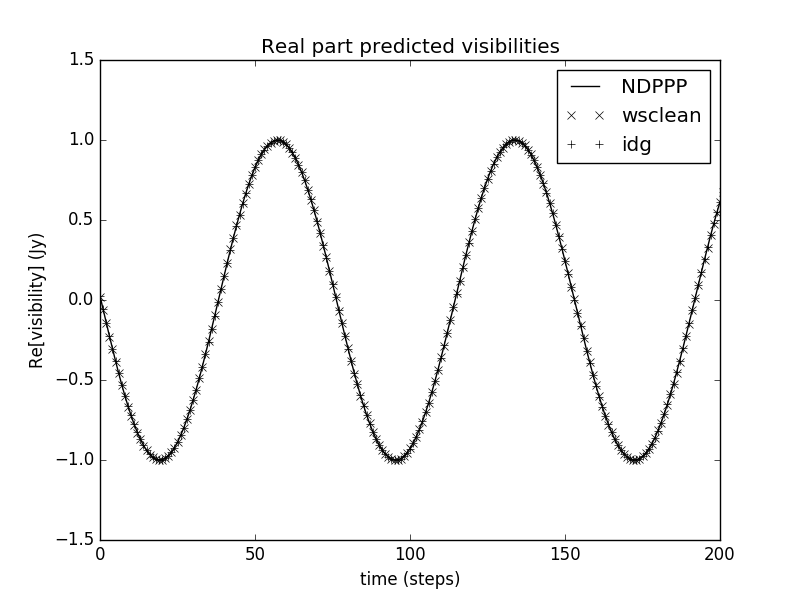}
  \includegraphics[width=0.33\textwidth]{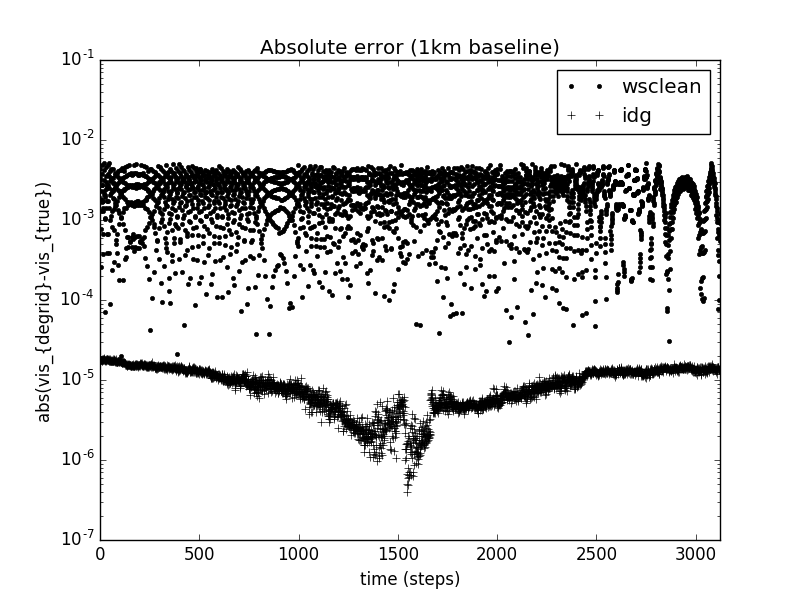}
  \includegraphics[width=0.33\textwidth]{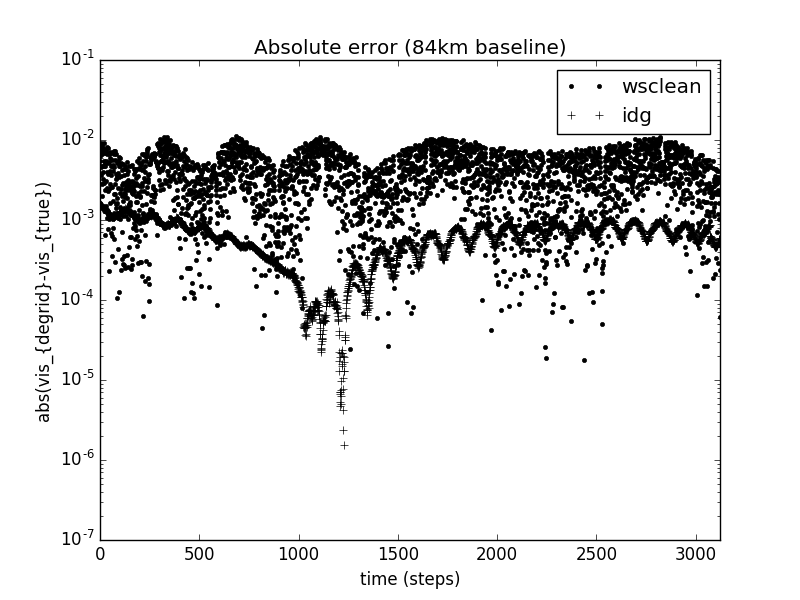}
   \caption{\textit{Left}: Real value of visibilities for a point source as predicted by direct evaluation of the ME, and degridding by the classical gridder and image domain gridder.
   The visibilities are too close together to distinguish in this graph. \textit{Middle, right}: Absolute value of the difference between direct evaluation and degridding
   for a short (1km) and a long (84km) baseline. On the short baseline the image domain gridder rms error of \SI{1.03e-05}{Jy} is about \num{242} times lower than the classical gridder rms error of \SI{2.51e-03}{Jy}.
   On the long baseline the image domain gridder rms error of \SI{7.10e-04}{Jy} is about seven times lower than the classical gridder error of \SI{4.78e-03}{Jy}.
   }
   \label{fig:residual_plot}
\end{figure*}

\begin{figure*}[!tbp]
  \includegraphics[width=0.45\textwidth]{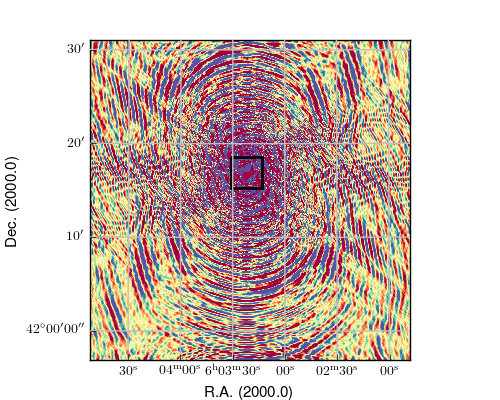}
  \includegraphics[width=0.45\textwidth]{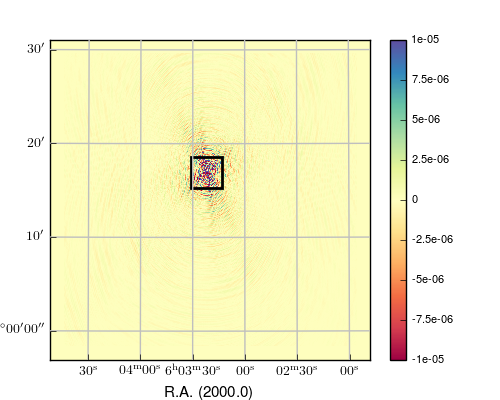}
  \caption{Residual image for the classical gridder in wsclean (left) and the image domain gridder (right).
  The color-scale for both images is the same, ranging from \SI{-1.0e-05}{Jy \per beam} to \SI{1.0e-05}{Jy \per beam}. The rms value of the
  area in the box centered on the source is about \num{19} times lower for image domain gridding (\SI{7.6e-06}{Jy \per beam}) than for classical gridding (\SI{1.3e-4}{Jy \per beam}).
  The rms value over the entire image is about \num{17} times lower for image domain gridding (\SI{1.1e-06}{Jy \per beam}) than for classical gridding (\SI{2.1e-05}{Jy \per beam})}
   \label{fig:residual_image}
\end{figure*}

A much larger difference can be seen in the residual visibilities in Figure \ref{fig:residual_plot} and the residual image in Figure \ref{fig:residual_image}.
The factor-18 lower noise in the residual image means an increase of the dynamic range limit  by that factor.
This increase will of course only be realized when the gridding errors are the limiting factor.

\begin{figure*}[!tbp]
  \includegraphics[width=0.45\textwidth]{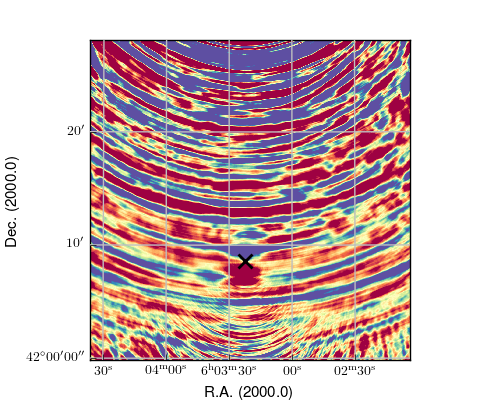}
  \includegraphics[width=0.45\textwidth]{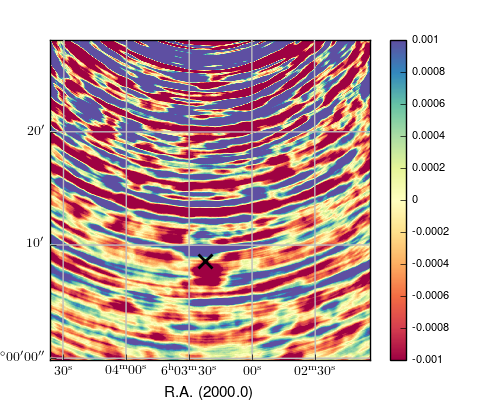}
  \caption{Image of simulated data of a source outside the field of view with classical gridding (left) and image domain gridding (right).
  The color-scale for both images is the same, ranging from \SI{-1.0e-03}{Jy \per beam} to \SI{1.0e-03}{Jy \per beam}.
  Position of the source is just outside the image to the north. The aliased position of the source within the image is indicated by an `X'.
  The image is the convolution of the PSF with the actual source and all its aliases.
  In the image for the classical gridder (left) the PSF around the alias is just visible. In the image for the image domain gridder (right) the alias is
  almost undetectable. The better alias suppression  has little effect on the overall rms value since this is dominated by the side lobes of the PSF around the actual source.
  The rms value over the imaged area is 2\% lower for image domain gridding (\SI{1.35}{Jy \per beam}) than for classical gridding (\SI{1.38e-03}{Jy \per beam}).
  }
  \label{fig:outside-fov}
\end{figure*}

In Figure \ref{fig:outside-fov} a modest 2\% better suppression of an outlier source is shown. This will have little impact on the dynamic range.

\subsection{Imaging observed data}

This imaging job was run on one of the GPU nodes of LOFAR Central Processing cluster.
This node has two Intel(R) Xeon(R) E5-2630 v3 CPUs running at \SI{2.40}{GHz}.
Each CPU has eight cores. With hyperthreading each core can run 2 threads simultaneously.
All in all, 32 threads can run in parallel on this node.

The node also has four NVIDIA Tesla K40c GPUs. Each GPU has a compute power of 4.29 Tflops (single precision).

The purpose of this experiment is to measure the run time of an imaging job large enough to make a reasonable extrapolation to a full-size job.
This is not a demonstration of the image quality that can be obtained, because that requires a more involved experiment.
For example, direction-dependent corrections are applied, but they were filled with identity matrices.
Their effect is seen in the runtime, but not in the image quality.

The dataset is again the ``toothbrush'' dataset used also for the simulations.
The settings are chosen to image the full field of LOFAR at the resolution for an observation including all remote stations (but not the international stations).
The image computed is 30000$\times$30000 pixels with 1.2asec/pixel.

After imaging 10\% was clipped on each side, resulting in a 24000 $\times$ 24000 pixel image, or 8$\deg$$\times$8$\deg$.
The weighting scheme used is Briggs' weighting, with the robustness parameter set to 0.
The cleaning threshold is set to 100 mJy, resulting in four iterations of the major cycle.
Each iteration takes about 20 minutes.

\begin{figure*}[!tbp]
  \includegraphics[width=\textwidth, trim = 105pt 370pt 90pt 0, clip=true]{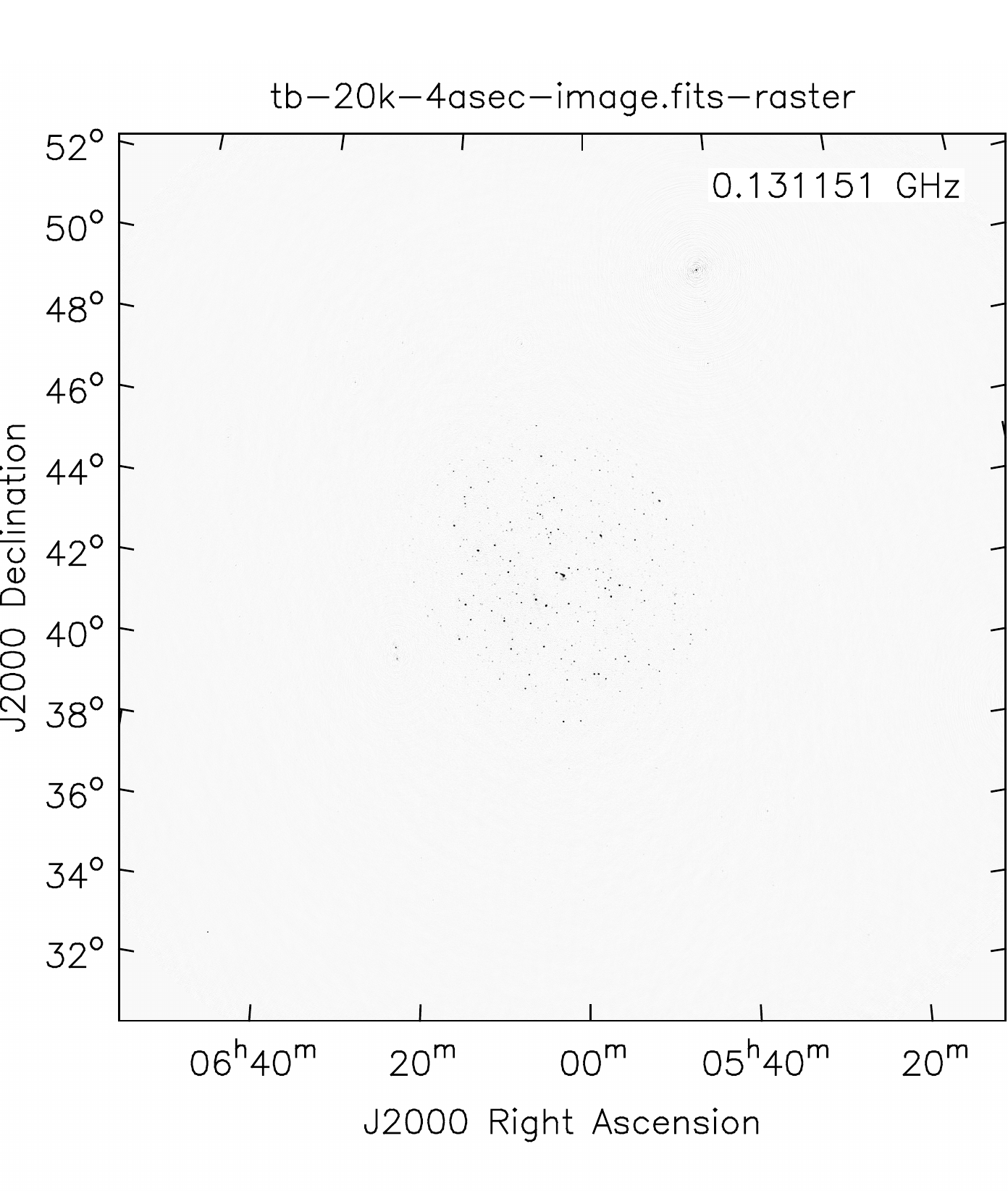}
  \caption{Large image (20000 $\times$ 20000 pixel) of the toothbrush field. The field of view is 22\textdegree $\times$ 22\textdegree at a resolution of \SI{4}{arcsec} per pixel.
  Cleaned down to \SI{100}{mJy} per beam, taking four major cycles.}
  \label{fig:toothbrush}
\end{figure*}

\section{Conclusions \& future work}

The image domain gridding algorithm is designed for the case where the cost of computing the gridding kernels is a significant part of the total cost of gridding.
It eliminates the need to compute a (sampled) convolution kernel by directly working in the image domain.
This not only eliminates the cost of computing a kernel, but is also more accurate compared to using an (over)sampled kernel.

Although the computational cost of the new algorithm is higher in pure operation count than classical gridding, in practice
it performs very well. On some (GPU) architectures it is even faster than classical gridding even when the cost of computing the convolution functions is not included.
This is a large step forward, since it is expected that for the square kilometer array (SKA), the cost of computing the convolution kernels will dominate the total cost of gridding.

Both in theory and simulation, it has been shown that image domain gridding is at least as accurate as classical gridding as long as a good taper is used.
The optimal taper has been derived.

The originally intended purpose of image domain gridding, fast application of time and direction dependent corrections,
has not yet been tested, as the corrections for the tests in this paper have been limited to identity matrices.
The next step is to use image domain gridding to apply actual corrections.

Another possible application of image domain gridding is calibration.
In calibration, a model is fitted to observed data. This involves the computation of residuals and derivatives.
These can be computed efficiently by image domain gridding whereby the free parameters are the A-term.
This would allow to fit directly for an A-term in the calibration step, using a full image as a  model.

\begin{appendix}
\section{Derivation of the optimal window}
The optimal window is derived by writing out the expression for the mean energy in the side lobes in terms of coefficients $a_k$.
This expression contains a double integral: one integral is over the extent of the side lobes, and one over all allowed positions in the sub-grid.
The double integral can be expressed in terms of special functions. The expression for the mean energy then reduces to a weighted vector norm,
where the entries of the weighting matrix are given in terms of the special functions.
The minimization problem can then readily be solved by singular value decomposition.

The square of the effective window given in \eqref{eq:effective_window} is
\begin{equation}
  \|f(x,s)\|^2 = \sum_{k=0}^{L-1} \sum_{l=0}^{L-1} a_k a_l e^\frac{j 2\pi (k-l) s}{L} \operatorname{sinc}(x - k)\operatorname{sinc}(x - l). \\
\end{equation}
This can be written as a matrix product:
\begin{equation}
  f^2(x,s) = \mathbf{a}^{H} \left( \mathbf{Q}(x) \odot \mathbf{S}(s) \right) \mathbf{a},
\end{equation}
where the elements of matrix $\mathbf{Q}(x)$ are given by $q_{i,j} = \operatorname{sinc}(x - k)\operatorname{sinc}(x - l)$ and
the elements of matrix $\mathbf{S}(s)$ are given by $s_{kl} = e^\frac{j 2\pi (k-l) s}{L}$.
The equation for the error \eqref{eq:error} can now be written as
\begin{equation}
  \varepsilon = \mathbf{a}^{H} \left( \mathbf{\overline{Q}} \odot \mathbf{\overline{S}} \right) \mathbf{a} = \mathbf{a}^{H} \left( \mathbf{\overline{R}} \right) \mathbf{a}
,\end{equation}
where $\mathbf{\overline{R}} = \mathbf{\overline{Q}} \odot \mathbf{\overline{S}}$ and
\begin{equation}
  \mathbf{\overline{Q}} = \int_{-\infty}^{0} \mathbf{Q}(x)\,\mathrm{d}x + \int_{L}^{\infty} \mathbf{R}(x)\,\mathrm{d}x
  \label{eq:meanR}
,\end{equation}
and
\begin{equation}
  \mathbf{\overline{S}} = \int_{-(L-\beta+1)/2}^{(L-\beta+1)/2} \mathbf{S}(s) ds
.\end{equation}
To evaluate the entries of $\mathbf{\overline{Q}}$ the following integral is needed:
\begin{equation}
  \begin{split}
    & \int \frac{\sin^2(\pi x)}{\pi^2(x^2 + kx)}dx =  \\
    & \frac{1}{2\pi^2k}\left(  \operatorname{Ci}(2\pi(k+x)) - \log(k+x) + \right. \\
    & \left. -\operatorname{Ci}(2\pi x)  + \log(x) \right), \quad \forall k \in \mathbb{Z}
  \end{split}
,\end{equation}
where $\operatorname{Ci}(x)$ is the \emph{cosine integral}, a special function defined by
\begin{equation}
  \operatorname{Ci}(x) = \int_x^\infty \frac{\cos t}{t}\,\mathrm{d}t
.\end{equation}
The entries of matrix $\mathbf{\bar{S}}$ are given by:
\begin{equation}
  \begin{split}
    \bar{s}_{kl} & = \frac{1}{L-\ +1}\int_{-(L-\beta+1)/2}^{(L-\beta+1)/2} e^\frac{j 2\pi (k-l) s}{L} ds  \\
    & = \frac{1}{L-\beta+1}\left[ -\frac{jL}{2\pi (k-l)} e^\frac{j2\pi (k-l) s}{L} \right]_{-(L-\beta+1)/2}^{(L-\beta+1)/2} \\
    & = \frac{L}{\pi (k-l)(L-\beta+1)} \sin(\pi (k-l) (L-\beta+1)/L) \\
    & = \operatorname{sinc}((k-l)(L-\beta+1)/N)
  \end{split}
.\end{equation}

\end{appendix}

\begin{acknowledgements}

This work was supported by the European Union, H2020 program, Astronomy ESFRI and Research Infrastructure Cluster (Grant Agreement number: 653477).
The first author would like to thank Sanjay Bhatnagar and others at NRAO, Soccoro, NM, US, for their hospitality and discussions on the A-projection algorithm in May 2011
that ultimately were the inspiration for the work presented in this paper. We would also like to thank A.-J. van der Veen for his thorough reading of and comments on a draft version.

\end{acknowledgements}

\bibliographystyle{aa}
\bibliography{references}

\end{document}